\documentclass[letterpaper]{article} 
\usepackage[draft]{aaai2026}  
\usepackage{times}  
\usepackage{helvet}  
\usepackage{courier}  
\usepackage[hyphens]{url}  
\usepackage{graphicx} 
\usepackage{natbib}  
\usepackage{caption} 
\usepackage{algorithm}
\usepackage{algorithmic}
\usepackage{amsmath}  
\usepackage{amssymb}

\usepackage{newfloat}
\usepackage{listings}
\DeclareCaptionStyle{ruled}{labelfont=normalfont,labelsep=colon,strut=off} 
\floatstyle{ruled}
\newfloat{listing}{tb}{lst}{}
\floatname{listing}{Listing}
\title{Amadeus: Autoregressive Model with Bidirectional Attribute Modelling for Symbolic Music}
\author{
    Hongju Su\textsuperscript{\rm 1}, 
    Ke Li\textsuperscript{\rm 1,2}*, 
    Lan Yang\textsuperscript{\rm 1,2}, 
    Honggang Zhang\textsuperscript{\rm 1}, 
    Yi-Zhe Song\textsuperscript{\rm 2}
}
\affiliations{
    \textsuperscript{\rm 1}School of Artificial Intelligence, Beijing University of Posts and Telecommunications, China \\
    \textsuperscript{\rm 2}SketchX, CVSSP, University of Surrey, United Kingdom \\
    hongjusu@bupt.edu.cn, like1990@bupt.edu.cn, ylan@bupt.edu.cn, zhhg@bupt.edu.cn, y.song@surrey.ac.uk \\
    *Corresponding Author \\
    https://github.com/lingyu123-su/Amadeus \\
}

\usepackage{bibentry}

\begin{document}

\maketitle

\begin{abstract}
Existing state-of-the-art symbolic music generation models predominantly adopt autoregressive or hierarchical autoregressive architectures, modelling symbolic music as a sequence of attribute tokens with unidirectional temporal dependencies, under the assumption of a fixed, strict dependency structure among these attributes. However, we observe that using different attributes as the initial token in these models leads to comparable performance. This suggests that the attributes of a musical note are, in essence, a concurrent and unordered set, rather than a temporally dependent sequence. Based on this insight, we introduce Amadeus, a novel symbolic music generation framework. Amadeus adopts a two-level architecture: an autoregressive model for note sequences and a bidirectional discrete diffusion model for attributes. To enhance performance, we propose Music                  Latent Space Discriminability Enhancement Strategy(MLSDES), incorporating contrastive learning constraints that amplify discriminability of intermediate music representations. The Conditional Information Enhancement Module (CIEM) simultaneously strengthens note latent vector representation via attention mechanisms, enabling more precise note decoding. We conduct extensive experiments on unconditional and text-conditioned generation tasks. Amadeus significantly outperforms SOTA models across multiple metrics while achieving at least 4$\times$ speed-up. Furthermore, we demonstrate training-free, fine-grained note attribute control feasibility using our model. To explore the upper performance bound of the Amadeus architecture, we compile the largest open-source symbolic music dataset to date, AMD (Amadeus MIDI Dataset), supporting both pre-training and fine-tuning. 
\end{abstract}


\section{Introduction}

\begin{figure}
\centering
\includegraphics[width=1\linewidth]{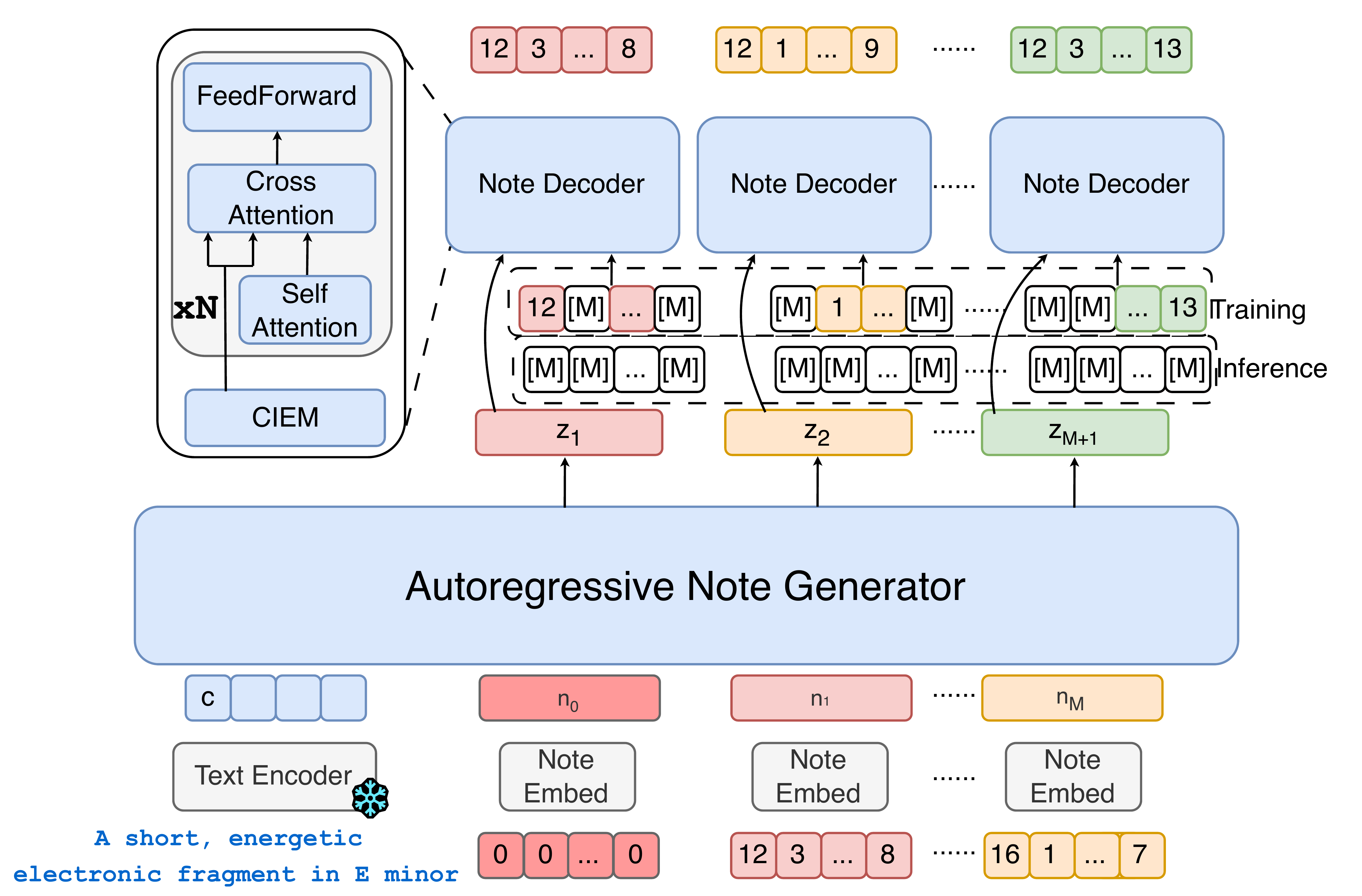}
\caption{Overview of the Amadeus framework. The framework consists of two main components: a note generator that auto regressively generates note-level latent variables, and a bidirectional discrete diffusion note decoder that decodes these latent variables into multi-dimensional attributes iteratively. }
\vspace{-0.3cm}
\label{fig:framework}
\end{figure}

Algorithmic composition has long been a core challenge in music theory, intersecting technical hurdles with music aesthetics\cite{xenakis1992formalized}, creativity research\cite{hiller1979experimental}, and human-computer interaction\cite{cope2000algorithmic}. In recent years, deep learning-based music generation methods~\cite{dhariwal2020jukebox, agostinelli2023musiclm, huang2018music, poptransformer2020,hsiao2021compound} have driven rapid advancements in this field. From a modelling perspective, existing approaches primarily fall into two categories: audio-based generation~\cite{dhariwal2020jukebox, agostinelli2023musiclm} and symbolic music-based generation~\cite{huang2018music, poptransformer2020,dong2023mmt,qu2024mupt}. Compared to audio-based methods, symbolic music modelling—which represents music as sequences of performance signals (\textit{e.g.}, MIDI)—offers significant advantages in lossless editing capability and compact representation.  

In sequential generation tasks, auto-regressive modelling~\cite{radford2018improving} has garnered widespread attention due to its high performance and scalability. Symbolic music can be characterized as a time series of notes, each composed of multiple attributes (\textit{e.g.}, pitch, instrument). Consequently, auto-regressive modelling is widely used for symbolic music generation~\cite{roy2025text2midi,musecoco2023,huang2018music,poptransformer2020} and has achieved notable success. However, ~\cite{roy2025text2midi} directly applying auto-regressive modelling to generate attribute sequences step-by-step (treating intra-note attributes as independent tokens) results in excessively long sequences (max sequence length = number of notes × number of attributes), making it difficult to model long-term musical structures. Some studies~\cite{ryu2024nmt} decouple the generation process into two steps: first auto regressively generating note-level latent variables, then decoding them into multi-dimensional attributes. While this approach effectively reduces note sequence length, it compromises the quality of the generated symbolic music.  

To address these limitations, we argue that while treating notes as time series for auto regressive modelling is valid, intra-note attributes (\textit{e.g.}, pitch, type, velocity, instrument and duration) are inherently concurrent and temporally independent sets (unordered), rather than sequences with intrinsic temporal dependencies. Thus, the theoretical basis for auto regressively modelling intra-note attributes is questionable—for instance, existing studies (\textit{e.g.}, CPWord~\cite{hsiao2021compound} using type to guide generation, NMT~\cite{ryu2024nmt} starting with metric or pitch as the initial attribute) achieve comparable generation quality, suggesting that strict temporal dependency assumptions may be unnecessary. Imposing sequential order and generating attributes one-by-one not only introduces efficiency bottlenecks but also restricts the potential for controllable symbolic
 music generation.  

To overcome these constraints, we propose Amadeus—a novel framework for symbolic music generation. Amadeus adopts a two-level architecture: an autoregressive model for the note sequence and a bidirectional discrete diffusion model at the attribute level. To enhance performance, we propose a Music Latent Space Discriminability Enhancement Strategy(MLSDES)—incorporating enhanced contrastive learning constraints to amplify the discriminability of intermediate music representations in the model. Simultaneously, to enhance note decoding accuracy, we propose the Conditional Information Enhancement Module (CIEM), which amplifies discriminative features in auto regressively generated note latent vectors through attention mechanisms and incorporates global contextual information, thereby guiding note decoding.

To validate Amadeus's modelling capability and controllability, we designed experiments across three dimensions: i) Unconditional symbolic music generation to verify foundational modelling capabilities; ii) Training-free fine-grained attribute control by leveraging attribute independence and specifiable decoding steps (specifying attribute values during decoding in the unconditional Amadeus model); iii) Text-controlled symbolic music generation. All experiments demonstrate that Amadeus significantly outperforms existing state-of-the-art methods in generation quality, condition fidelity, attribute controllability, and inference speed. Notably, the method allows flexible trade-offs between quality and speed via hyperparameter tuning. Finally, we compiled, curated, and open-sourced the largest symbolic music dataset to date: AMD (Amadeus MIDI Dataset). This dataset includes a 1.9-million-sample pre-training set and a 320,000-sample high-quality fine-tuning set with textual annotations, enabling in-depth exploration of data scale and model parameter size impacts on Amadeus's performance.  

Our core contributions are summarized as follows:  
\begin{itemize}
    \item We propose the Amadeus architecture—performing autoregressive modelling at the note level and bidirectional modelling at the attribute level via discrete diffusion—enhanced by the MLSDES and CIEM modules to boost representational capacity. This approach delivers robust symbolic music modelling and exceptional controllability.
    \item Through comprehensive and fair comparisons in unconditional and controllable symbolic music generation tasks, we empirically demonstrate Amadeus's superiority over existing methods in music quality, condition adherence, attribute controllability, and inference speed.
    \item We compile and open-source the largest public symbolic music dataset, AMD, comprising 1.9 million pre-training samples and 320,000 annotated fine-tuning samples.
\end{itemize}   
  
\section{Related Work}

\subsection{Computer-Aided Composition} The earliest attempt at computer-aided music composition was the \emph{Illiac Suite} (later known as \emph{String Quartet No. 4}), created using the ILLIAC I computer~\cite{hiller1979experimental}. Its fourth movement employed probabilistic models and Markov chains to generate rhythm and melody. Subsequent research~\cite{cope2000algorithmic} explored rule-based and probabilistic systems, with methods like transition matrices built on existing melodies~\cite{collins2011chopin}. These approaches introduced mathematical formalism into music composition but suffered from poor musical quality due to limited expressive modelling capacity.

Later efforts turned to machine learning and neural networks~\cite{nakamura2015autoregressive,mogren2016crnngan,Raffel2016LearningBasedMF}, which improved fluency but still lacked convincing musicality. With deep learning, Transformer-based models~\cite{huang2018music} significantly enhanced generation performance, yet struggled with modelling long symbolic sequences. \cite{poptransformer2020}introduced REMI tokenization and used Transformer-XL\cite{dai-etal-2019-transformer} to extended context but introduced training complexity and unstable results.

Attribute-merging methods~\cite{hsiao2021compound} reduced sequence length by combining note features, but led to degraded expressiveness. Improvements such as attribute-level teacher-forcing~\cite{ryu2024nmt} enhanced model capacity but imposed strict sequential assumptions among unordered attributes, limiting generation diversity, speed, and controllability.

\subsection{Autoregressive Model Improvements} GPT-style autoregressive models are constrained by slow generation, weak long-range dependency modelling, and lack of bidirectionality. One line of work proposed partially autoregressive generation:~\cite{yu2023megabyte} replaced tokenization with letter patches and used a two-stage autoregressive process to model mega-length sequences. Despite flexibility, generation remained dependent on sequential decoding. Patch-based improvements~\cite{meta_blt} slightly boosted performance but retained these limitations.

Another line of work moved beyond autoregression. Discrete diffusion models~\cite{austin2021structured} enabled parallel generation by discretizing continuous diffusion processes. Follow-up studies~\cite{nie2025scalingmaskeddiffusionmodels,nie2025large} demonstrated scalability, but training remained slow and inference unstable.

Many symbolic music models~\cite{wang2025notagenadvancingmusicalitysymbolic,d3pmforsymbolicmusic,guo2025moonbeammidifoundationmodel,wu2024melodyt5unifiedscoretoscoretransformer} adopt these methods, yet ignore key domain properties. For instance,~\cite{ryu2024nmt} treat note attributes as ordered, violating their inherent independence; others~\cite{d3pmforsymbolicmusic} neglect the temporal structure of note sequences. Such mismatches limit generation quality, control, and efficiency.

\section{Methodology}
\subsection{Preliminary}
In symbolic music generation tasks, a musical work $\mathcal{M}$ is represented as a time series $\mathcal{M} = \{\mathcal{N}_0,\mathcal{N}_1, \mathcal{N}_2, \dots, \mathcal{N}_M\}$, where $M+1$ denotes the number of notes, $\mathcal{N}_0$ is an all-zero vector, serving as the initial note for all musical work $\mathcal{M}$. Each note $\mathcal{N}_M \in \mathbb{R}^K$ is defined by a $K$-dimensional feature vector, with each dimension corresponding to a core musical attribute (detailed attribute definitions are available in the supplementary material). The joint state of these attributes uniquely determines each note in the musical sequence.  

\textbf{Note Embedding}
The note $\mathcal{N}_m$ has each attribute $\mathcal{N}_m^{k}$ treated as a token. The note embedding module maps every attribute $\mathcal{N}_m^{k}$ to a $d$-dimensional embedding vector $\mathbf{e}_m^{k} \in \mathbb{R}^d$. The composite note embedding $\mathbf{n}_m \in \mathbb{R}^d$ is then computed by aggregating all attribute embedding vectors $\mathbf{e}_m^{k}$ with a positional embedding vector through summation:
\begin{equation}
\mathbf{n}_m = \sum_{k=0}^{K-1} \mathbf{e}_m^{k} + \mathbf{p}_m
\end{equation}
where $\mathbf{p}_m \in \mathbb{R}^d$ is a learnable absolute positional embedding vector corresponding to the $m$-th note.  

\textbf{Note Generator}
Composite note embedding sequence $\{\mathbf{n}_0, \mathbf{n}_1,\dots, \mathbf{n}_m\}$ is fed into the note generator $G$, which operates as a Transformer-based decoder. Conditioned on the input sequence $\mathbf{c}$, it generates the latent vector $\mathbf{z}_{m+1}$ for the next note in an autoregressive manner:  
\begin{equation}  
\mathbf{z}_{m+1} = G(\mathbf{c}, \mathbf{n}_0, \mathbf{n}_1, \dots, \mathbf{n}_m)  
\end{equation}

\textbf{Note Decoder}
The note decoder in symbolic music generation methods predominantly relies on sequential autoregressive decoding of note attributes~\cite{ryu2024nmt}. However, intrinsic properties of notes (such as pitch, duration, velocity, etc.) are fundamentally concurrent and atemporal features. This decoding paradigm not only violates their essential nature but also introduces significant efficiency bottlenecks by enforcing strictly sequential generation of all attributes. To address these challenges, we propose a Discrete Diffusion Model ~\cite{nie2025large}(DDM) based note decoding framework operating at the attribute granularity. This framework significantly outperforms existing autoregressive approaches~\cite{musecoco2023,bhandari2025text2midi} in both generation quality and inference speed.  

\textbf{Discrete Diffusion Model (DDM)}
We employ Masked Diffusion Model (MDM)—a discrete diffusion model—to model bidirectional dependencies between note attributes, enabling parallel attribute decoding. In MDM, each note $\mathbf{n}_{m+1}$ is treated as the starting state $x_0$. During the forward process, the noise level at each time step  $t$ is controlled by the hyper-parameter  $\alpha_t = 1 - t$ . With probability $p_{\text{mask}}$, each token is independently replaced by the mask symbol $[M]$. The forward process is defined as:  
\begin{equation}
\nonumber
q_{t|0}(\boldsymbol{x}_t|\boldsymbol{x}_0) = \prod_{k = 0}^{K - 1} q_{t|0}(x_t^k|x_0^k)
\text{,} 
\end{equation}
\begin{equation}
\quad q_{t|0}(x_t^k|x_0^k) =
\begin{cases}
\alpha_t, & x_t^k = x_0^k \\
1 - \alpha_t, & x_t^k = \texttt{[M]}
\end{cases}
\end{equation}

The reverse process is initiated from a fully masked sequence $\boldsymbol{x}_T$ and progressively recovers the masked tokens:  
\begin{equation}
\nonumber
q_{s|t}(\boldsymbol{x}_s|\boldsymbol{x}_t) = \prod_{k = 0}^{K - 1} q_{s|t}(\boldsymbol{x}_s^k|\boldsymbol{x}_t)
\text{,} 
\end{equation}

\begin{equation}
q_{s|t}(\boldsymbol{x}_s^k|\boldsymbol{x}_t) =
\begin{cases}
1, & x_t^k \neq \texttt{[M]},\ x_s^k = x_t^k \\
\frac{s}{t}, & x_t^k = \texttt{[M]},\ x_s^k = \texttt{[M]} \\
\frac{t - s}{t} q_{0|t}(\boldsymbol{x}_s^k|\boldsymbol{x}_t), & x_t^k = \texttt{[M]},\ x_s^k \neq \texttt{[M]} \\
0, & \text{otherwise}
\end{cases}
\end{equation} 

\subsection{Music Latent Space Discriminability Enhancement Strategies}
To enhance the quality of generated music and reduce the learning difficulty for subsequent note decoders, we proposes a Music Latent Space Discriminability Enhancement Strategies (MLSDES) based on enhanced contrastive learning~\cite{wang2025diffuse}. This strategy introduces an improved contrastive loss function at the intermediate representation layer of the note generator model (rather than at the final bottleneck layer), which amplifies the distance between latent vectors of different music samples to strengthen semantic discriminability in the latent space. The loss function is designed as follows:  
\begin{equation}  
\mathcal{L}_{\mathrm{CL}} = -\log \frac{1}{N(N-1)} \sum_{i=1}^N \sum_{j \neq i} \exp\!\left( -\frac{1 - \cos(\mathbf{h}_i, \mathbf{h}_j)}{\tau} \right)  
\end{equation}  
where $\mathbf{h}_i,\mathbf{h}_j \in \mathbb{R}^d$ denote the latent vectors of music samples, obtained by applying average pooling to the intermediate-layer latent vectors of their constituent notes. \(\cos(\cdot)\) represents the cosine distance, $N$ is the number of music samples in a batch, and \(\tau > 0\) is a temperature hyper-parameter.  

\begin{figure*}[ht]
\centering
\includegraphics[width=0.9\textwidth]{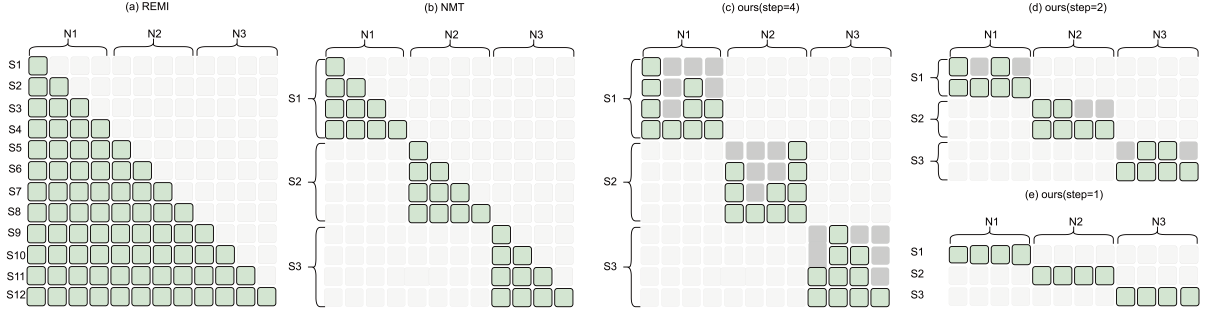}
\caption{Comparative schematics of note decoding processes: (a) REMI; (b) NMT; (c-e) Our step-adjustable note decoding (step=4, 2, 1).}
\vspace{-0.2cm}
\label{fig:inference}
\end{figure*}

\subsection{DDM Based Note Decoder}
\textbf{Conditional Information Enhancement Module (CIEM)} The CIEM is constructed based on the cross-attention mechanism within the Transformer decoder (structure shown in Fig.~\ref{fig:framework}). This module reconstructs the output latent vector $\mathbf{z}_{m+1}$ from the note generator through multi-head attention mechanism, generating an enhanced representation $\hat{\mathbf{z}}_{m+1}$ infused with global structural information. This enhanced representation then serves as the control condition for the subsequent note decoder (a discrete diffusion model). Compared to direct use of the original latent vector $\mathbf{z}_{m+1}$, the enhanced representation $\hat{\mathbf{z}}_{m+1}$ enhances discriminative features in $\mathbf{z}_{m+1}$ through Self-Attention (SA) mechanism and incorporates global musical context from $\mathbf{z}_1$ into the current prediction through Cross-Attention (CA) mechanism. The computational process is formally defined as:  
\begin{equation}  
\hat{\mathbf{z}}_{m+1} = \text{CA}\Big({\text{SA}(\mathbf{z}_{m+1})}, \mathbf{z}_1, \mathbf{z}_1 \Big)  
\end{equation}  
where \(\text{SA}(\cdot)\) denotes self-attention mechanism and \(\text{CA}(\cdot)\) represents the cross-attention mechanism.

\textbf{Step-Adjustable Note Decoding}
The forward process of our note decoding framework adheres to the standard pipeline of discrete diffusion models~\cite{nie2025large}. Its core operation involves applying time-step-dependent independent masking operations to each note attribute (treated as an independent token) in the sequence. The reverse process initiates from a partially masked sequence (during training) or a fully masked sequence (during inference). After the note latent vector $\mathbf{z}_m$ generated by the note generator is processed by the Conditional Information Enhancement Module (CIEM) to obtain the enhanced conditional signal $\hat{\mathbf{z}}_m$, this signal serves as a conditioning input to guide the progressive recovery of all masked attributes via cross-attention mechanisms. At each time-step $t$ (total steps $T$), the system precisely computes the number of tokens $\mathtt{num_{\text{tk}}}$ to decode using an approximate uniform recovery schedule:
\begin{equation}  
\mathtt{num_{\text{tk}}}=
\begin{cases}
\left\lfloor \frac{\mathtt{num_{m}}}{T} \right\rfloor + 1, & \text{if } t < \mathtt{num_{m}} \bmod T \\
\left\lfloor \frac{\mathtt{num_{m}}}{T} \right\rfloor, & \text{otherwise}
\end{cases}
\end{equation}  
where $\mathtt{num_{m}}$ is the total number of masked tokens in note.
The model then predicts the attribute distribution for all tokens, selects the $\mathtt{num_{\text{tk}}}$ tokens with the highest confidence scores to recover and fix their attributes, while re-masking all remaining tokens.  

Fig.~\ref{fig:inference} visually illustrates the flexible step-adjustable decoding mechanism of our method during the reverse process. Using a sequence containing three notes (each with four attributes) as an example (the horizontal axis denotes token sequence length; the vertical axis indicates decoding iterations—higher rows correspond to more iterations/longer latency): REMI~\cite{poptransformer2020} and NMT~\cite{ryu2024nmt} prevent parallel attribute decoding by enforcing sequential constraints. Each attribute is decoded one at a time, requiring exactly 12 iterations for 3 notes and 4 attributes per note. In contrast, our method eliminates inter-attribute sequential dependencies, leveraging discrete diffusion to enable attribute-parallel decoding while permitting flexible control over total steps $T$ (where $T$ also adjusts $\mathtt{num_{\text{tk}}}$). By setting different $T$ values: $T=4$ (more steps), $T=2$, or $T=1$ (fewest steps), overall decoding latency can be drastically altered, thereby achieving an effective trade-off between decoding speed and quality.

\textbf{Note Decoder Loss Function}
We adopts a masked weighted cross-entropy loss function $\mathcal{L}_{\text{CE}}$ as the optimization objective, defined as follows:  
\begin{equation}  
\mathcal{L}_{\text{CE}} = -\sum_{k=1}^K \frac{1}{p_{\text{mask}}^k} \cdot m_k \cdot \sum_{v=1}^{V_k} \mathbb{I}[\mathcal{N}^{k}_{m+1} = v] \log p(v \mid \mathcal{N}_{m+1}^{k})  
\end{equation}  
where $p_{\text{mask}}^k$ denotes the probability of applying a mask to the $k$-th attribute during the forward diffusion process. This probability acts as a weighting factor to balance the frequency differences in masking across attributes. $m_k \in \{0,1\}$ is a binary mask indicator—when $m_k=1$, it indicates that the corresponding position is non-padded and the attribute is masked. $V_k$ represents the vocabulary size of the $k$-th attribute. The indicator function $\mathbb{I}[\cdot]$ outputs 1 if the ground-truth value of the $k$-th attribute at the $m+1$-th note equals $v$, otherwise 0. The probability term $p(v \mid n_{m+1}^{(k)})$ signifies the model's predicted probability. Through the synergistic mechanism of inverse probability weighting ($1/p_{\text{mask}}^k$) and mask filtering ($m_k$), this loss function focuses exclusively on optimizing prediction errors at effective masked attribute positions.

\subsection{Training Objective}
The overall training objective is a combination of the discriminability enhancement constraint loss $\mathcal{L}_{\mathrm{CL}}$ and the note decoder cross-entropy loss $\mathcal{L}_{\mathrm{CE}}$:
\begin{equation}
\mathcal{L} = \lambda \mathcal{L}_{\mathrm{CL}} + \mathcal{L}_{\mathrm{CE}}
\end{equation}
where \( \lambda \) is a hyper-parameter to balance the two terms.

\section{Experiment}

To ensure a fair validation of our method's superiority, all comparative experiments were conducted under conditions of comparable parameter counts (170M) and datasets. Specifically, i) In the unconditional symbolic music generation task, we trained and tested on the SOD~\cite{leopold_crestel_2017_1416204} dataset and Lakh Clean~\cite{Raffel2016LearningBasedMF} dataset respectively, comparing against four representative methods (\textit{i.e.},CPWord~\cite{hsiao2021compound}, MMT~\cite{dong2023mmt}, NMT~\cite{ryu2024nmt}, and REMI~\cite{poptransformer2020}); ii) In the training-free note attribute control task, we validated Amadeus's capability to generate high-quality symbolic music while specifying note attributes, using an unconditionally trained model based on the Lakh Clean dataset; iii) In the text-conditioned symbolic music generation task, we compared against three representative methods (\textit{i.e.},Text2Midi~\cite{bhandari2025text2midi},MuseCoco~\cite{musecoco2023} and T2M-inferalign~\cite{roy2025text2midi}) on the text-annotated MidiCaps~\cite{Melechovsky2024} dataset. Notably, to explore the performance ceiling of our model, we constructed and open-sourced the AMD dataset (containing 1.9 million pre-training samples and 320,000 text-annotated fine-tuning samples). On this dataset, we trained an expanded model Amadeus-M with 500M parameters.

\begin{table}[t]
\centering
\vspace{-0.1cm}
\setlength{\tabcolsep}{1mm}
\begin{tabular}{lccc|ccc}
\hline
\multicolumn{4}{c|}{LakhClean} & \multicolumn{3}{c}{SOD} \\
\hline
Model & SC $\uparrow$ & PE $\downarrow$ & PCE $\downarrow$ & SC $\uparrow$ & PE $\downarrow$ & PCE $\downarrow$ \\
\hline
CPWord & 0.93 & 4.19 & 2.64 & 0.90 & 4.29 & 2.89 \\
MMT & 0.94 & 4.09 & 2.61 & 0.91 & 4.38 & 2.90 \\
NMT & 0.91 & 4.36 & 2.82 & 0.91 & 4.32 & 2.93 \\
REMI & 0.95 & 3.45 & 2.38 & 0.90 & 4.60 & 3.05 \\
\hline
\textbf{Amadeus} & \textbf{0.97} & \textbf{2.20} & \textbf{1.97} & \textbf{0.92} & \textbf{4.24} & \textbf{2.82} \\
\hline
\end{tabular}
\caption{Unconditional music generation results on the LakhClean dataset and SOD dataset.}
\label{tab:uncond-lakh}
\vspace{-0.1cm}
\end{table}

\subsection{AMD Dataset}
\textbf{Pre-training Dataset} : By integrating GigaMIDI~\cite{Lee_2025}, AriaMIDI~\cite{bradshaw2025ariamididatasetpianomidi}, SymphonyNet~\cite{liu2022symphonygenerationpermutationinvariant}, MidiCaps(excluding its test set), XMIDI~\cite{xmusic2025}, and 80,000 self-crawled high-quality symbolic music segments, we constructed a pre-training corpus after data cleaning. This corpus comprises 1.9 million music files and approximately 4 billion music events (equivalent to 32 billion attribute tokens). To our knowledge, this pre-training dataset is currently the largest open-source symbolic music pre-training corpus.

\textbf{Supervised Fine-tuning Dataset}: We integrated MidiCaps, XMIDI and 80,000 self-crawled high-quality symbolic music segments as raw materials. After data cleaning and annotation augmentation, we ultimately constructed a high-quality supervised fine-tuning dataset containing 320,000 samples. All samples include structured textual descriptions. More detailed information on the dataset is available in the Supplementary Materials.
\subsection{Metrics} 
For the unconditional generation task and training-free note attribute controlled task, our evaluation focuses on structural consistency (Scale Consistency, SC) and quality (Pitch Entropy, PE,Pitch-Class Entropy, PCE) of generated symbolic music used in~\cite{dong2018pypianoroll,DBLP:conf/ismir/WuY20}. For the text-conditioned generation task, The evaluation system comprises: i) CLAP Score~\cite{laionclap2023,evans2024stableaudioopen}, measuring the similarity between the prompt text and generated audio; and ii) a fine-grained music characteristics metric set, which comprehensively validates the alignment of generated symbolic music with text-specified requirements across five dimensions: Tempo Bin with Tolerance(TBT), Correct Key (CK)~\cite{melechovsky2024mustango}, Correct Time Signature (CTS), Coverage of Instruments (CI), and Coverage of top3 moods (CM$_{\text{top3}}$). Detailed definitions of all metrics are provided in the supplementary material.

\begin{table}[t]
\centering
\vspace{-0.1cm}
\begin{tabular}{lc|ccc}
\hline
Controlling Attribute & Value & SC$\uparrow$ & PE$\downarrow$  & PCE$\downarrow$  \\
\hline
Amadeus & - & 0.97 & \textbf{2.20} & 1.97 \\
\hline
Instrument & Piano & 0.96 & 3.21 & 2.15 \\
Tempo      & 153   & 0.96 & 2.91 & 2.05 \\
Chord      & C7    & \textbf{0.98} & 2.92 & \textbf{1.81} \\
Velocity   & 60    & 0.96 & 2.98 & 2.09 \\
\hline
\end{tabular}
\caption{The results of Training-free Note Attribute Control, which were achieved on an unconditional symbolic music generation model trained on the Lakh Clean dataset.}
\label{tab:attr-control}
\vspace{-0.1cm}
\end{table}

\begin{table*}[ht]
\centering
\vspace{-0.1cm}
\begin{tabular}{l|c|cccccc}
\hline
 & Speed(notes/s)  & CLAP$\uparrow$  & TBT$\uparrow$  & CK$\uparrow$  & CTS$\uparrow$  & CI$\uparrow$  & CM$_{\text{top3}}$$\uparrow$  \\
\hline
Text2Midi    & 4.02   & 0.19 & 31.76 & 22.22 & 84.15 & 19.92 & 60.57 \\
MuseCoco     & 1.67 & 0.19 & 34.21 & 14.66 & 94.24 & 22.42 &  38.18 \\
T2M-inferalign           & 4.02 & 0.20 & 39.32 & 29.80 & 84.32 & 20.13 & 47.74 \\
\hline
Amadeus   & 16.23 & 0.20 & 73.93 & 39.31 & 96.98 & 26.01 & 65.52 \\
Amadeus-M    & 10.51 & \textbf{0.21} & \textbf{76.31} & \textbf{43.07} & \textbf{97.02} & \textbf{27.11} & \textbf{66.39} \\
\hline
\end{tabular}
\caption{Text-conditioned music generation results. We evaluated the text-condition fidelity, fine-grained note attribute accuracy, and generation speed of the generated music on the MidiCaps dataset.}
\label{tab:text-control}
\vspace{-0.2cm}
\end{table*}

\subsection{Unconditional Symbolic Music Generation}
The unconditional symbolic music generation results shown in Tab.~\ref{tab:uncond-lakh} demonstrates that:
i) The model achieves optimal performance across all evaluation metrics on both datasets—demonstrating significant superiority on the multi-genre Lakh Clean dataset while also exhibiting exceptional performance on the SOD dataset, which primarily features orchestral and classical music. Given that orchestral and classical music demand stronger fine-grained attribute modelling and long-range structural modelling capabilities, these results fully confirm that our model significantly outperforms traditional autoregressive and other baseline models in capturing the intrinsic development of symbolic music; 
ii) The highest SC score indicates that the model possesses more stable structural coherence when generating long-sequence music; 
iii) Significant advantages in PE and PCE further corroborate the model's precise modelling ability for harmonic complexity, with the generated music surpassing baseline methods in both harmonic richness and quality.

\subsection{Training-free Note Attribute Control }
Music can be characterized as a sequence of notes, each containing multiple attributes. The ability to directly specify specific note attributes during generation offers a novel perspective for controllable symbolic music generation. Our approach eliminates temporal dependencies between attributes, enabling controllable decoding order.Users can prioritize specifying and decoding specific attributes, with subsequent attributes generated based on these choices (\textit{e.g.}, prioritizing the instrument attribute ensures the generated music effectively utilizes that instrument). Tab.~\ref{tab:attr-control} presents evaluation results for music generated under constraints specifying the Instrument, Tempo, Chord, and Velocity attributes. Results show negligible degradation compared to unconditional Amadeus, validating our training-free method’s ability in controlling fine-grained attributes without compromising symbolic music generation quality.This capability greatly expands the application scenarios for symbolic music generation, proving particularly valuable for real-time accompaniment generation and complex orchestration tasks.

\subsection{Text-Conditioned Symbolic Music Generation}
To validate the comprehensive performance of our method on text-conditioned symbolic music generation tasks, we conducted systematic experiments on the MidiCaps dataset with fine-grained textual annotations (results shown in Tab.~\ref{tab:text-control}). The results demonstrate that our method achieves breakthroughs across three critical dimensions—generation quality, control precision, and inference speed—outperforming baseline models in both text-condition fidelity (CLAP Score) and fine-grained metrics, while accelerating inference to 16.23 note/s (tested on an RTX3090 24G GPU): a 4× speed-up over existing approaches. This comprehensive superiority validates the technical excellence of our method for efficient and controllable symbolic music generation.

Regarding  control capability, our model exhibits precise modelling across multiple musical metrics: improvements in TBT and CK reach 88.02\% and 31.91\%, respectively, proving its robust rhythmic modelling capabilities. A 2.90\% improvement in CTS confirms that the attribute-level bidirectional modelling mechanism effectively captures intrinsic harmonic functional relationships. A 16.01\% enhancement in CI reflects the model's refined semantic modelling ability, while an 8.17\% increase in CM$_{\text{top3}}$ highlights its exceptional emotion modelling performance. These quantitative results collectively demonstrate that our method achieves precise modelling and decoding of note attributes through the DDM-Based Note Decoder.

\begin{figure*}
\centering
\includegraphics[width=0.9\linewidth]{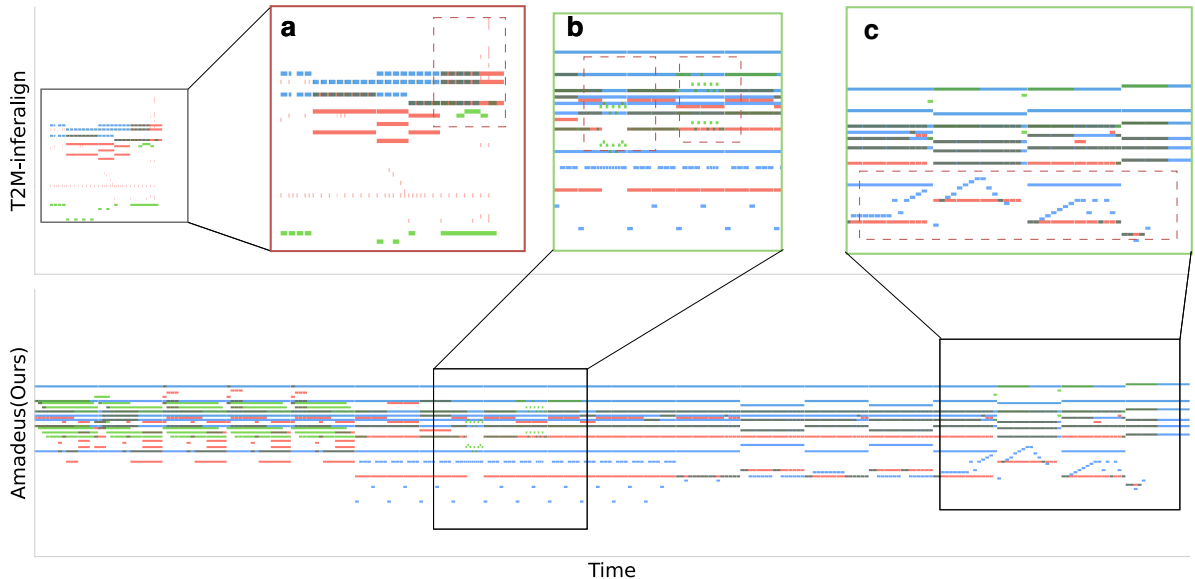}
\caption{Comparison of piano roll visualizations of MIDI files generated by T2M-inferalign (upper panel) and our method (lower panel).}
\label{fig:compare}
\vspace{-0.1cm}
\end{figure*}

Fig.~\ref{fig:compare} presents qualitative analysis results for the text-conditioned symbolic music generation task. The text prompt states: "This lengthy electronic piece, infused with a touch of easy listening, is a melodic and relaxing journey. The acoustic guitar takes the lead, accompanied by a string ensemble, piano, shana, and synth strings. Set in A minor and maintaining a 4/4 time signature, it moves at an Allegro tempo of 120 beats per minute. Throughout the composition, the chords E and A alternate frequently, creating a meditative atmosphere with a hint of Christmas spirit and a motivational undertone." The figure displays piano roll visualizations of MIDI files generated by T2M-inferalign (upper panel) and our method (lower panel), respectively. The horizontal axis represents time, each coloured block denotes a note whose length corresponds to its duration and vertical position to its pitch (detailed track-wise visualization is provided in the appendix).
Based on a traditional autoregressive architecture for note decoding, T2M-inferalign is constrained by limited long-sequence modelling capability—resulting in fewer generated notes, low note density per time-step, and impoverished variations in pitch-duration combinations. This indicates its inadequacy in generating complex harmonies. Notably, the scattered tomato-coloured short notes in Fig.~\ref{fig:compare} (a)'s red dashed box further reveal the model’s difficulty in producing sustained, musically coherent note sequences.
In contrast, the music generated by our method exhibits significantly higher textural complexity and harmonic sophistication. Within the red dashed box in Fig.~\ref{fig:compare} (b), the main development demonstrates a rhythmic evolution: transitioning from dense long notes to regular green short notes before returning to long notes. This pattern highlights our model’s ability to generate intricate rhythmic structures, thereby enriching the auditory experience. The blue melodic line within the red dashed box in Fig.~\ref{fig:compare} (c) precisely aligns with the prompt’s requirement for "melodic". This not only validates our model’s exceptional adherence to textual instructions but also underscores its advanced generative quality in melody modelling.

\begin{table*}[ht]
\centering
\vspace{-0.1cm}
\begin{tabular}{ccc|c|c|ccccc}
\hline
MLSDES & CIEM & Steps & Speed(notes/s) & CLAP$\uparrow$  & TBT$\uparrow$  & CK$\uparrow$  & CTS$\uparrow$  & CI$\uparrow$  & CM$_{\text{top3}}$$\uparrow$  \\
\hline
$\times$ & $\times$ & 8 & 16.21 & 0.16 & 69.78 & 36.62 & 94.17 & 23.60 & 61.72 \\
$\surd$  & $\times$ & 8 & 16.24 & 0.17 & 70.01 & 38.76 & 94.33 & 23.79 &  62.35\\
$\times$  & $\surd$ & 8  & 16.23 & 0.20 & 70.03 & 38.20 & 95.15 & 24.01 & 62.60 \\
\hline
$\surd$  & $\surd$  & 1 & 32.12 & 0.19 & 59.84 & 28.19 & 89.30 & 20.08 & 37.93 \\
$\surd$  & $\surd$  & 4 & 20.11  & 0.20 & 66.46 & 39.00 & \textbf{97.38} & 25.80 & \textbf{67.40} \\
$\surd$  & $\surd$  & 8  & 16.23 & \textbf{0.20} & \textbf{73.93} & \textbf{39.31} & 96.98 & \textbf{26.01} & 65.52 \\
\hline
\end{tabular}
\caption{Ablation study of MLSDES, CIEM and denoising steps on text-conditioned symbolic
 music generation.}
\label{tab:ablation-denoise-rep-text}
\vspace{-0.2cm}
\end{table*}

\subsection{Ablation Study}
Amadeus achieves significant improvements in both symbolic
music generation quality and speed. To deeply investigate the impact of MLSDES and CIEM on generation quality and
the trade-off role of the step hyperparameter in Step-Adjustable Note Decoding between speed and quality,
we designed systematic ablation experiments. As shown in the Tab. 4: 
i) CIEM significantly improves the quality of generated symbolic music. This is because CIEM introduces global information into the DDM based note decoder, which greatly ensures the continuity of the generated symbolic music sequence. Without CIEM, even with MLSDES, the overall quality of the generated audio remains poor. This proves that the CIEM module is indispensable.
ii) After introducing the MLSDES, model performance improves
across different situation. This is because the strategy enhances the discriminability of note representations, significantly reducing the learning diﬀiculty of the note decoding module and thus recovering note attributes more accurately; 
iii) As the number of denoising steps decreases, the generation speed significantly
increases (up to 32.12 notes/s), but the model performance generally exhibits a declining trend. The iterative parallel decoding process enables the model to fully
integrate currently decoded information and gradually
optimize generation results—increasing steps enhances
quality at the cost of speed; 
iv) Higher steps do not
always correspond to better performance. For example, when MLSDES and CIEM are enabled, the model achieves superior results on CTS and CM$_{\text{top3}}$metrics when step is 4.
The core reason is that the confidence-based retention mechanism at each step resembles a local greedy search.
Moderately reducing steps helps the model escape local optima to achieve globally superior solutions; 
v) Whether pursuing the highest quality (where Amadeus operates at 16.23 notes/s, significantly outperforming the comparative methods’ 4.02 notes/s) or prioritizing maximum speed (where generation quality is lowest but still comparable to the comparative methods), Amadeus demonstrates comprehensive superiority.
\subsection{Effects of Dataset Scale and Parameter Scale}
We adopted a two-stage training strategy: pre-training on the AMD pre-training dataset, followed by supervised fine-tuning on the fine-tuning dataset. To further enhance model capability, we scaled up the model size, denoted as Amadeus-M (500M parameters). As shown in the Tab.~\ref{tab:text-control}, Amadeus-M achieves significant improvements across all performance metrics in the text-controlled music generation task. Notably, the increased model parameter count resulted in a decreased generation speed of 10.51 notes/s, yet this speed remained significantly higher than the 4.02 notes/s of other comparative methods. 
Experiments verified that its performance can be further enhanced with continuously increasing data volume and parameter count.

\section{Conclusion}
We propose Amadeus, a novel framework for symbolic music generation. Amadeus adopts a two-level architecture: an autoregressive model for modelling the note sequence and a bidirectional discrete diffusion model at the attribute level. To further enhance performance, we introduce Music Latent Space Discriminability Enhancement Strategies (MLSDES) and Conditional Information Enhancement Module (CIEM). We conduct extensive experiments on both unconditional and text-conditioned generation tasks. Across all benchmarks, Amadeus significantly outperforms existing state-of-the-art models while achieving at least 4× speed-up. However, synergistic optimization potential between computational efficiency and generation  quality remains in the current framework. Future work will focus on exploring lossless acceleration techniques and real-time generation.

\appendix

\section{MIDI Protocol, Tokenization Method, and Note Attributes}
The Musical Instrument Digital Interface (MIDI) is an industry-standard electronic communication protocol that defines a comprehensive set of codes for musical notes and performance actions, enabling electronic instruments, computers, mobile devices, and other stage equipment to connect, synchronize, and exchange performance data in real time. MIDI provides a complete framework for describing the performance states of different instruments and plays a crucial role in modern music composition workflows. Due to the flexibility of MIDI signals, artists often use Digital Audio Workstations (DAWs) to rapidly iterate on musical ideas by editing MIDI data and assigning virtual instruments. Once the desired effect is achieved, the final version may be recorded with real instruments. This workflow significantly lowers the barrier to music production and democratizes music creation.

There are various methods for converting MIDI signals into tokens. We recommend interested readers refer to the \texttt{miditok} library, which provides a detailed introduction to specific tokenization processes. In our work, we adopt an improved note-based encoding proposed by NMT. Specifically, each note is represented by the following attributes:
\begin{itemize}
    \item \textbf{Type}: each representing a different combination of metrical changes or continuations.
    \item \textbf{Beat}: The relative position of the note within the same measure.
    \item \textbf{Chord}: The chord to which the current note belongs.
    \item \textbf{Tempo}: The playback speed of the note; generally, a higher tempo indicates a faster song.
    \item \textbf{Instrument}: The instrument performing the current note.
    \item \textbf{Pitch}: The pitch of the note, represented by 128 discrete values according to the MIDI specification.
    \item \textbf{Duration}: The duration for which the note is played.
    \item \textbf{Velocity}: The intensity with which the note is played, determining its loudness.
\end{itemize}
These attributes uniquely specify a note. Importantly, each attribute describes an independent aspect of the note, and the set of attributes is inherently free from unidirectional sequential dependencies. Technically, the order of these attributes can be arbitrarily permuted without affecting the representation of the note itself, which clearly goes beyond the scope of unidirectional dependency. As observed in prior work, different methods may use different attributes (such as \texttt{type} or \texttt{pitch}) as the first token, yet achieve comparable performance. This further demonstrates the insufficiency of assuming unidirectional dependencies among note attributes.

\section{Mutual Information Analysis of Token Attributes}

To assess the dependency structure among the attributes, we conducted a mutual information (MI) and conditional entropy analysis. This empirical study is intended to inform the modeling strategy of musical event sequences, particularly the suitability of autoregressive vs. bidirectional generation paradigms.

\subsection{Experimental Setup}

We tokenized  corpus of LakhClean dataset MIDI files into notes, from which we extracted the following attribute dimensions: 
\texttt{Beat Position}, \texttt{Pitch}, \texttt{Velocity}, \texttt{Duration}, \texttt{Instrument}, \texttt{Chord}, \texttt{Tempo}, and \texttt{Type}.

The total amount of note tokens in the dataset is 82985704.
For each attribute, we applied label encoding and computed the empirical mutual information between all attribute pairs:
\[
I(X; Y) = \sum_{x,y} p(x, y) \log \frac{p(x, y)}{p(x)p(y)}
\]
We further normalized MI using the geometric mean of marginal entropies to obtain a symmetric measure in $[0,1]$:
\[
\text{NMI}(X; Y) = \frac{I(X; Y)}{\sqrt{H(X)H(Y)}}
\]

\subsection{Results and Observations}

\begin{table*}[h]
\centering
\begin{tabular}{lcccccccc}
\hline
         & Beat   & Pitch  & Velocity & Duration & Instrument & Chord  & Tempo  & Type \\
\hline
Beat     & 1.1703 & 0.5128 & 0.5128   & 0.5128   & 0.5128  & 0.0008 & 0.4905 & 0.0964 \\
Pitch    & 0.5128 & 3.5981 & 0.5308   & 0.5401   & 0.8679  & 0.0007 & 0.4205 & 0.0314 \\
Velocity & 0.5128 & 0.5308 & 1.6784   & 0.5222   & 0.5356  & 0.0007 & 0.4205 & 0.0314 \\
Duration & 0.5128 & 0.5401 & 0.5222   & 2.4884   & 0.5961  & 0.0007 & 0.4205 & 0.0314 \\
Instrument  & 0.5128 & 0.8679 & 0.5356   & 0.5961   & 2.6845  & 0.0007 & 0.4205 & 0.0314 \\
Chord    & 0.0008 & 0.0007 & 0.0007   & 0.0007   & 0.0007  & 0.0046 & 0.0008 & 0.0000 \\
Tempo    & 0.4905 & 0.4205 & 0.4205   & 0.4205   & 0.4205  & 0.0008 & 1.0186 & 0.0042 \\
Type     & 0.0964 & 0.0314 & 0.0314   & 0.0314   & 0.0314  & 0.0000 & 0.0042 & 0.1040 \\
\hline
\end{tabular}
\caption{Raw Mutual Information (nats)}
\end{table*}

\begin{table*}[h]
\centering
\begin{tabular}{lcccccccc}
\hline
         & Beat   & Pitch  & Velocity & Duration & Instrument & Chord  & Tempo  & Type \\
\hline
Beat     & 1.0000 & 0.2499 & 0.3659   & 0.3005   & 0.2893  & 0.0114 & 0.4492 & 0.2764 \\
Pitch    & 0.2499 & 1.0000 & 0.2160   & 0.1805   & 0.2793  & 0.0056 & 0.2197 & 0.0513 \\
Velocity & 0.3659 & 0.2160 & 1.0000   & 0.2555   & 0.2523  & 0.0082 & 0.3216 & 0.0752 \\
Duration & 0.3005 & 0.1805 & 0.2555   & 1.0000   & 0.2306  & 0.0068 & 0.2641 & 0.0617 \\
Instrument  & 0.2893 & 0.2793 & 0.2523   & 0.2306   & 1.0000  & 0.0065 & 0.2543 & 0.0594 \\
Chord    & 0.0114 & 0.0056 & 0.0082   & 0.0068   & 0.0065  & 1.0000 & 0.0120 & 0.0004 \\
Tempo    & 0.4492 & 0.2197 & 0.3216   & 0.2641   & 0.2543  & 0.0120 & 1.0000 & 0.0128 \\
Type     & 0.2764 & 0.0513 & 0.0752   & 0.0617   & 0.0594  & 0.0004 & 0.0128 & 1.0000 \\
\hline
\end{tabular}
\caption{Normalized Mutual Information}
\end{table*}

The normalized mutual information matrix $\text{NMI}(X;Y)$ reveals several trends. First, dependencies among \texttt{Pitch}, \texttt{Velocity}, \texttt{Duration}, and \texttt{Instrument} are generally weak to moderate (for example, $\text{NMI}(\texttt{Pitch}, \texttt{Instrument}) \approx 0.28$). Such complex dependencies can be explained by basic musical knowledge: different instruments have distinct pitch ranges and playing styles, which influence the distribution of these attributes. The strongest dependency is observed between \texttt{Beat Position} and \texttt{Tempo} ($\text{NMI} \approx 0.45$), reflecting the rhythmic regularity commonly found in MIDI data.

We further analyzed asymmetric conditional entropy to evaluate directional predictability:
\[
H(Y|X) = H(Y) - I(X;Y)
\]
The estimated values:
\[
H(\texttt{Duration}|\texttt{Pitch}) \approx 1.95,\quad H(\texttt{Pitch}|\texttt{Duration}) \approx 3.06
\]
indicate that knowing \texttt{Pitch} reduces uncertainty in \texttt{Duration} more effectively than the reverse, suggesting a directional statistical influence from \texttt{Pitch} to \texttt{Duration}.

\subsection{Implications for Modeling}

The results indicate that the dependencies among note attributes are generally weak and not strictly unidirectional, as evidenced by the low maximum NMI values and the absence of any attribute that strongly determines the others. Furthermore, the observed dependencies are often mutual rather than hierarchical, suggesting that imposing a fixed generation order, as in conventional autoregressive (AR) models, may not be optimal. These findings support the adoption of bidirectional or non-autoregressive modeling strategies, which are better suited to capture the symmetrical and overlapping relationships among token attributes in symbolic music data.

\section{Training Details}
In our training setup, we employ gradient clipping with a threshold of 1 and utilize the AdamW optimizer in conjunction with a cosine learning rate scheduler. For unconditional generation, the maximum learning rate is set to $1\times 10^{-4}$; for conditional generation and large-scale pretraining, it is $2\times 10^{-4}$; and for text-guided fine-tuning, it is $5\times 10^{-5}$. The maximum number of training iterations is 100{,}000 for both unconditional and conditional generation tasks, and 300{,}000 for pretraining. Notably, we do not use the final checkpoint from pretraining, but rather select an intermediate checkpoint for subsequent experiments. Training is accelerated using distributed data parallelism (DDP) and the \texttt{accelerate} library with mixed-precision computation. For conditional generation, pretraining, and fine-tuning, we further employ gradient accumulation with a step size of 4.

\section{Data Collection and Processing}
\subsection{Pretraining Data}

Our pretraining dataset mainly combines the GigaMIDI, AriaMIDI, SymphonyNet corpora. The total number of files is 1.9 million, comprising about 4 billion events (approximately 32 billion attribute tokens). To our knowledge this is the largest open-source symbolic music pretraining dataset. To enhance data quality, we filter out GigaMIDI files that contain only drum tracks. Additionally, due to the relatively poor quality of some pretraining data, we also incorporate the MIDI files from the fine-tuning datasets into the pretraining collection. We do not include the MidiCaps test split in training to ensure fairness.

We follow a processing procedure partly inspired by NMT: first, we apply rule-based filtering to the MIDI files and detect attributes such as chords, time signatures, note durations, and instruments. We convert these attributes into event sequences, compile a vocabulary of tokens for each attribute, and then convert the event sequences into index files corresponding to this vocabulary. Notably, for any file missing tempo or time signature information, we set the default tempo to 120\,BPM and the default time signature to 4/4, in accordance with the MIDI specification.

\subsection{Supervised Fine-Tuning Data}

For the supervised (fine-tuning) dataset with text annotations, the processing differs from that of the pretraining data. We use the MidiCaps and XMIDI datasets, as well as an in-house dataset, as the raw corpora. XMIDI contains around 100k symbolic music pieces annotated by experts with emotion and genre labels, and our in-house dataset includes roughly 80k high-quality symbolic music pieces.

The overall processing pipeline is divided into two stages: text annotation and conversion to index sequences.

In the text annotation stage, we largely follow the MidiCaps methodology. We first extract information from the MIDI sequences directly using tools such as Music21 and Mido, capturing key (mode), time signature, tempo, durations, and instruments. We then synthesize each MIDI file into audio and extract features such as genre, mood, and chords. Based on the extracted information, we design prompt templates with examples and use DeepSeek-v3 to convert these prompts into textual annotations. For XMIDI, we simply use its provided emotion and genre labels as annotations.

In the second stage, we convert each piece into index files similarly to the pretraining data, with one major difference: we do not perform any instrument merging or trimming. Instrument merging (e.g., combining distorted guitar, overdrive guitar, and clean electric guitar into a single ``electric guitar'' category) is sometimes used to reduce sequence length, but it can severely harm the alignment between the generated music and the text control signal in the conditional generation task. Therefore, following the approach of \texttt{text2midi}, we assign each instrument to its own track to avoid merging. After this two-stage processing, we obtain a high-quality supervised symbolic music dataset of 320k examples with text annotations.

\section{Evaluation Metrics}
\subsection{Unconditional Generation}

For evaluation, we use metrics from both the audio domain and the symbolic domain to comprehensively assess model performance. In the unconditional generation task, for each model we generate 500 samples of a fixed length of 1024 tokens, render them to WAV audio, and uniformly trim each audio clip to 30 seconds.

We introduce a set of symbolic-domain metrics to comprehensively evaluate the quality and consistency of generated symbolic music. These include:                                 

\begin{itemize}
    \item \texttt{scale\_consistency}: Measures the maximum proportion of notes that fit within any assumed musical scale, reflecting the overall tonal coherence of the generated piece.
    \item \texttt{pitch\_entropy}: The Shannon entropy of the pitch histogram, quantifying the diversity and concentration of pitch usage. Lower entropy indicates that pitches are concentrated on a few core notes, suggesting clearer tonality and higher generation quality.
    \item \texttt{pitch\_class\_entropy}: The Shannon entropy of the pitch-class histogram, further characterizing the stability of pitch usage across pitch classes.
\end{itemize}

We compute these metrics using the methods provided by MusPy.

\subsection{Text-Conditioned Generation}

We use the MidiCaps test set of 500 text annotations to guide the generation of symbolic music, render the outputs to WAV, and trim them to 10 seconds (again generating 1024 tokens per sample). We evaluate from two perspectives: overall similarity between the prompt text and the generated audio, and accuracy of the conditioned attributes.

\paragraph{Overall Text–Audio Similarity} We use the CLAP score, which measures the cosine similarity between text and audio embeddings, bigger
CLAP score indicates better alignment between the text and audio. We follow the implementation of stable-audio.

\paragraph{Control Accuracy Metrics} The text annotation in MidiCaps includes attributes such as key, time signature, tempo, and instruments. Following previous works, We evaluate the accuracy of these attributes in the generated symbolic music using the following metrics:
\begin{itemize}
  \item \textbf{Tempo Bin with Tolerance (TBT)}: the ratio of samples where the predicted tempo falls within a tolerance range of 10 BPM from the true tempo. This metric accounts for slight variations in tempo that may still be musically acceptable. 
  \item \textbf{Correct Key (CK)}: the ratio of samples where the predicted key matches the true key, ignoring the octave. This metric evaluates the model's ability to capture the tonal center of the music. By convention, an undefined key is treated as C major.

  \item \textbf{Correct Time Signature (CTS)}: represents the ratio of generated MIDI files where the predicted time signature matches the true time signature. This metric evaluates the model's ability to capture the rhythmic structure of the music.

  \item \textbf{Coverage of Instruments (CI)} is the fraction where predicted instruments fully cover the ground truth; CI$_{\rm top1}$ is the fraction where at least one true instrument appears in the prediction.

    \item \textbf{Coverage of Mood (CM)}: the fraction of samples where the predicted mood matches the true mood. This metric evaluates the model's ability to capture the emotional content of the music;CM$_{\rm top3}$ is the fraction where at least three true moods appear in the prediction.(all mood if the number of moods is less than 3)

\end{itemize}

\section{Visualization of Generated Music}
To provide a visual representation of the generated symbolic music, we use the \texttt{pypianoroll} library to render the MIDI files into sheet music. Below are examples of generated tracks from both Amadeus and T2M-inferalign, showcasing the diversity and complexity of the outputs.
\begin{figure*}
\centering
\includegraphics[width=0.8\textwidth]{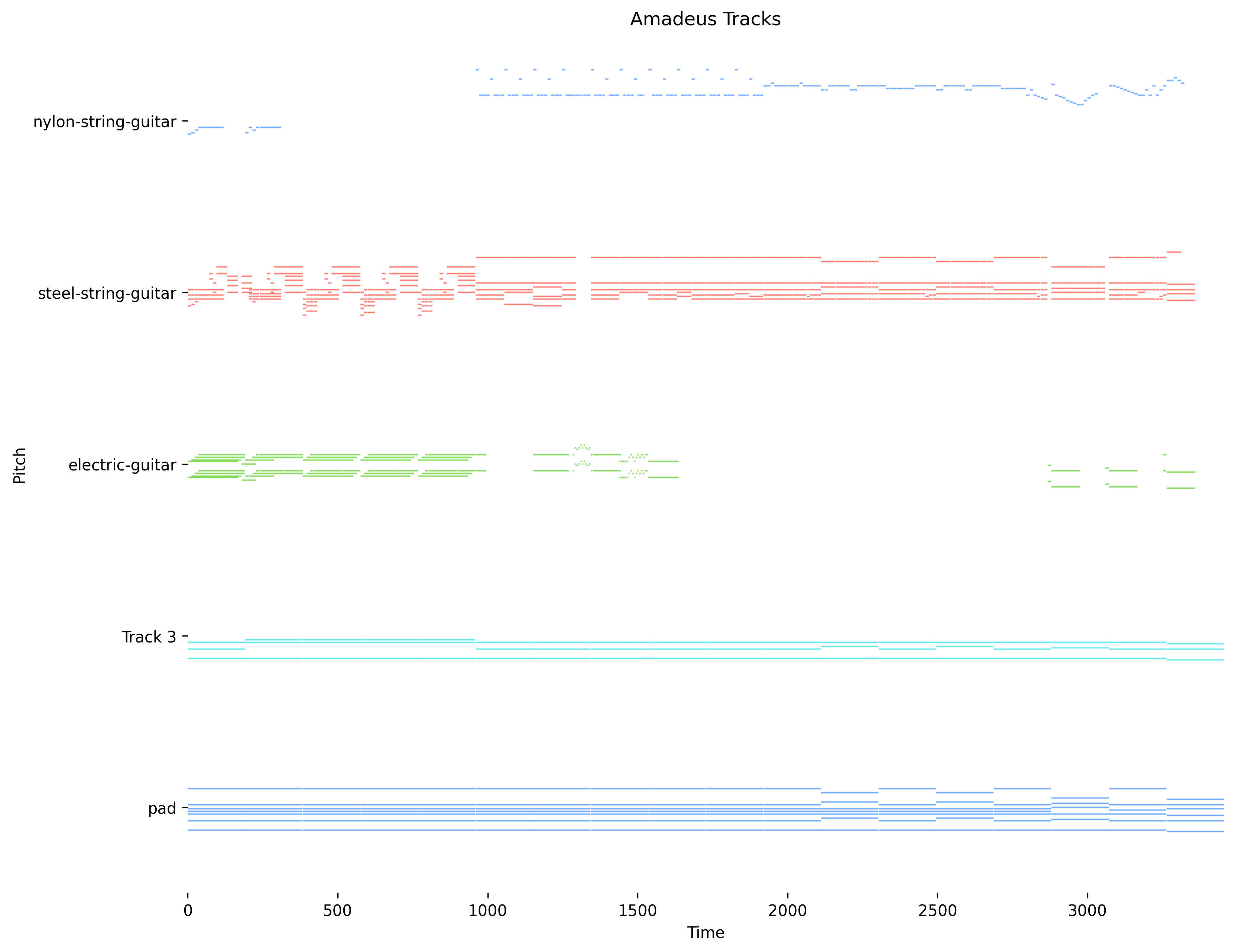}
\caption{Visualization of generated symbolic music by Amadeus.}
\end{figure*}

\begin{figure*}
\centering
\includegraphics[width=0.8\textwidth]{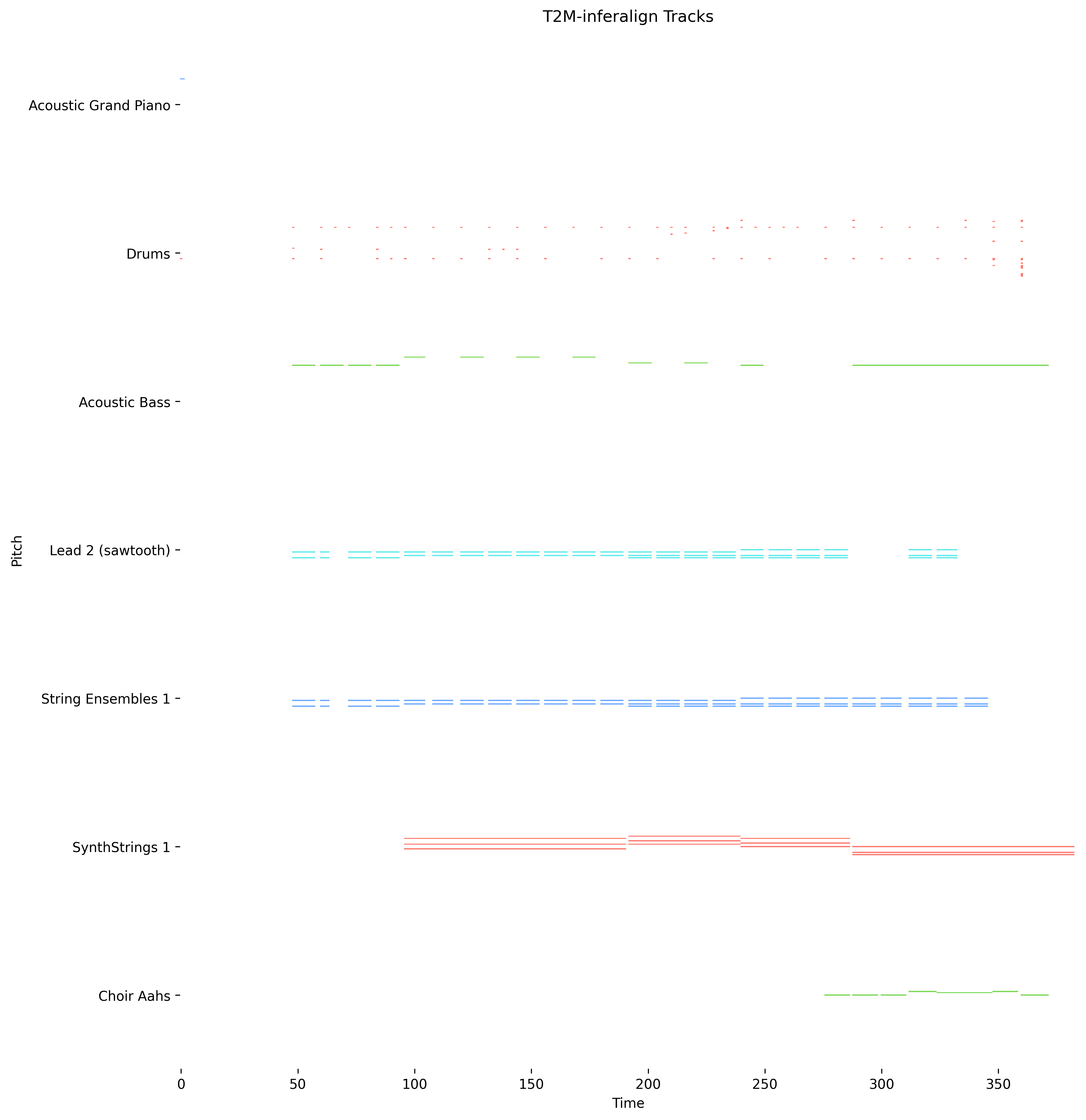}
\caption{Visualization of generated symbolic music by T2M-inferalign.}
\end{figure*}
\section{Limitations}

Despite the significant improvements achieved by our model over current state-of-the-art systems, several limitations remain. Due to constraints in the training dataset, our model demonstrates limited support for underrepresented musical genres such as traditional folk music, world music, and regional music styles. In addition, we did not adopt certain advanced architectural components such as Rotary Position Embedding (RoPE) and RMSNorm, which may have constrained the model’s overall performance. Furthermore, we did not conduct comprehensive inference optimization, resulting in suboptimal generation speed. Lastly, due to hardware limitations, we were unable to systematically study model scaling behavior, which may limit the reliability of scaling law predictions.

\section{Environmental Impact}

To reduce carbon emissions and make better use of computational resources, we employed mixed-precision training with the \texttt{accelerate} library. Compared to standard Distributed Data Parallel (DDP) training under the same power consumption, this setup achieved nearly 40\% faster training. This efficiency gain indirectly mitigates the environmental impact. We plan to explore additional energy-efficient training strategies in future work to further reduce environmental costs.

\section{Ethics Statement}

This research involves the use of human-composed music data for training and evaluation of a generative model. All music data were obtained from publicly available datasets or licensed sources and were used solely for academic research under fair use or equivalent scholarly exemptions. No personally identifiable information was involved. We did not conduct any experiments involving human subjects or listeners. All experiments were carried out in compliance with institutional guidelines on ethical research using existing data.

\section{Adverse Impact Statement}

This work presents a generative music model trained on datasets containing copyrighted music. Although the model is developed purely for academic research, we acknowledge the potential risks associated with unauthorized reproduction of copyrighted styles or melodies. To mitigate such risks, we refrain from releasing any trained models or audio outputs that may resemble specific copyrighted works. Our publication focuses on methodological contributions rather than commercial deployment. We explicitly discourage any misuse of our approach for content piracy, unauthorized distribution, or infringement of intellectual property rights.

\bibliography{aaai2026}
\end{document}